\documentclass[twoside]{article}
\usepackage{fleqn,espcrc2}
 \usepackage{epsfig}

\def\Journal#1#2#3#4{{#1} {\bf #2}, #3 (#4)}

\def\NPB{{\em Nucl. Phys.} B}

\def\PLB{{\em Phys. Lett.}  B}
\def\PRL{\em Phys. Rev. Lett.}
\def\PRD{{\em Phys. Rev.} D}

\def\slashchar#1{\setbox0=\hbox{$#1$}\dimen0=\wd0% 
\setbox1=\hbox{/}\dimen1=\wd1% 
\ifdim\dimen0>\dimen1%                        
\rlap{\hbox to 
\dimen0{\hfil/\hfil}}#1\else                                        
\rlap{\hbox to \dimen1{\hfil$#1$\hfil}}/\fi} 
\def\vepp{\varepsilon'}
\def\vep{\varepsilon}

\def\be{\begin{equation}}
\def\ee{\end{equation}}
\def\ba{\begin{eqnarray}}
\def\ea{\end{eqnarray}}

\def\cL{{\cal L}}
\def\cQ{{\cal Q}}
\def\cA{{\cal A}}

\title{Final State Interactions and $\vepp / \vep$}
\author{ E. Pallante
\address{Facultat de F\'{\i}sica, Universitat de Barcelona, Diagonal
647,\\ E-08028 Barcelona, Spain  }\thanks{SISSA, Via Beirut 2-4, 34014 Trieste,
 Italy, from September 2000.} }

\begin{document}
%\vspace*{4cm}

\begin{abstract}
I discuss the r$\hat{\mbox{o}}$le of strong final state interactions (FSI) in 
$K\to 2\pi$ decays. 
In this case  strong FSI effects can be resummed \cite{SHORT,FUTURE_PP} 
by solving the Omn\`es problem for $K\to 2\pi$ amplitudes.
Implications for the CP conserving $\Delta I=1/2$ ratio and the direct 
CP violation parameter $\vepp / \vep$ are also discussed.
\vspace{1pc}
\end{abstract}
\maketitle

\section{Introduction}

Strong final state interaction (FSI) effects are very important in the 
phenomenology of K meson decays. In the non--leptonic two--body 
$K\to \pi\pi$ decay the dominant FSI contribution is given by the 
elastic (soft) rescattering of the two pions in the final state. 

\begin{figure}[htb]
\centerline{\mbox{\epsfxsize=8cm\epsffile{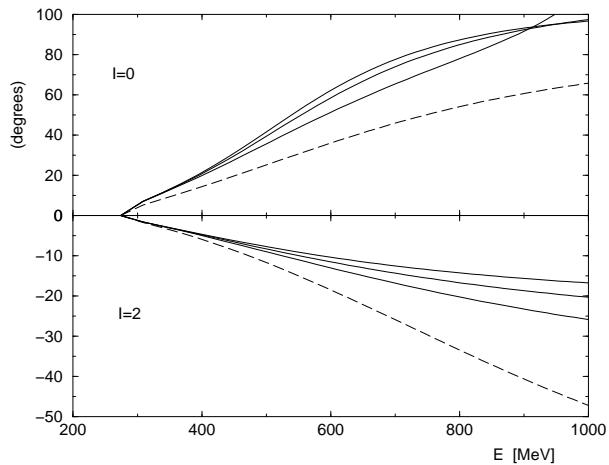}}}
\caption{
Phase shifts $\delta_0^{0;2}(s)$ with $I=0$ and 2, according to a 
fit \protect\cite{SCHENK} of experimental data
and used in the numerical analyses \protect\cite{SHORT,FUTURE_PP}.
Solid lines enclose the range covered by the experimental data, while
dashed lines show the unitarized lowest--order ChPT prediction.}
\label{fig:PHASESHIFT}
\end{figure}

At centre--of--mass energies around the kaon
mass, the strong S--wave $\pi$--$\pi$ scattering generates a large
phase shift difference
$\left(\delta^0_0 - \delta^2_0\right)(m_K^2)= 45^\circ\pm 6^\circ$
between the $I=0$ and $I=2$ partial waves \cite{GM}. 
This effect is taken into account by factoring out those phases
in the usual decomposition of $K\to \pi\pi$ amplitudes with definite isospin 
$I=0$ and $I=2$:
\be\label{eq:AIdef} 
{\cal A}_I \,\equiv\, A\left[ K\to (\pi\pi )_I\right]
\,\equiv\, A_I \; e^{i\delta^I_0} \; .
\ee
The presence of such a large phase shift difference also signals 
a large dispersive FSI effect in the moduli of the isospin amplitudes, since 
their imaginary and real parts are related by analyticity and unitarity. 
Intuitively, the behaviour of the $I=0$ and $I=2$ S--wave phase shifts
as a function of the total energy of the two pions as reported in  
 Figure~\ref{fig:PHASESHIFT}
suggests a large enhancement of the $I=0$
amplitude and a tiny suppression of the $I=2$ amplitude. 
%%%%%%%%%%%%%%%%%%%%%%%%%%%
%%%%%%%%%%%%%%%%%%%%%%%%%%%%%%%%%%%
The numerical estimate of this dispersive FSI effect is a difficult task, 
since it is dominated by long--distance (soft) contributions and reduces to 
a non--perturbative problem. 
  
The size of the FSI effect can be roughly 
estimated at one loop in Chiral Perturbation Theory (ChPT), where the 
rescattering of the two pions in the final state produces an enhancement 
of about $40\%$ in the $A_0$ amplitude \cite{KA91,BPP,JHEP}.
However, the fact that the one loop calculation still underestimates 
the observed $\delta_0^0$ phase shift indicates that a further enhancement 
should be produced by higher orders.
It is then necessary to resum FSI effects.

Lattice determinations of $K\to \pi\pi$ amplitudes could in principle take 
into account {\em automatically} strong FSI effects. 
However, the direct measurement 
of $K\to \pi\pi$ amplitudes on the lattice is still afflicted by a series of 
problems. More recently, a possible solution to those problems which overcomes 
the Maiani--Testa theorem \cite{MT} has been proposed \cite{LL}. 
In the meanwhile, most of the attempts up to date are based on the so called
 {\em indirect} method, i.e.
a two--steps procedure where first the simpler $K\to \pi$ matrix element
is measured on the lattice and second, the physical $K\to 2\pi$ matrix elements
are obtained by using a lowest--order ChPT relation  
between $K\to 0$, $K\to \pi$ and $K\to 2\pi$ \cite{Betal}. 
Neither the first nor the second step include FSI effects. Recently, the 
inclusion of one--loop ChPT contributions has been investigated in this 
context \cite{GP}.

Approaches based on effective low--energy models \cite{Trieste}
or the $1/N_c$ expansion \cite{Dortmund}
do include some one-loop corrections and find larger values for the
$A_0$ amplitude. However, the drawback in these cases might be 
the possible model dependence of the matching procedure with 
short--distance. 

Here, I discuss an approach to FSI effects in $K\to\pi\pi$ decays 
that has been recently proposed 
\cite{SHORT,FUTURE_PP}. It is based on the Omn\`es solution \cite{ALLOMNES}
for $K\to \pi\pi$ amplitudes, 
which permits the resummation of strong FSI effects to all orders in ChPT.
Intuitively, what the Omn\`es solution does is to correct a local weak 
$K\to \pi\pi$ transition with an infinite chain of pion--loop bubbles, 
incorporating the strong $\pi\pi\to\pi\pi$ rescattering to all orders 
in ChPT. 

A few properties of strong FSI effects can be useful in order to 
understand how they enter the prediction of $K\to\pi\pi$ decays:
\begin{itemize}

\item[$\bullet$ ] the production of the two pions and their subsequent
rescattering are two independent processes. 
 The rescattering process only depends on the quantum numbers 
(total isospin $I$ and total angular momentum $J$) of the two pions
in the final state and on their total energy.

\item[$\bullet$ ] At  centre--of--mass energies of the order of the K 
meson mass the elastic (soft) rescattering of the two pions is the dominant 
strong FSI effect. Hence, strong FSI effects in $K\to\pi\pi$ amplitudes
can be treated in a fully non--perturbative way. This implies that  
in the usual description of weak $\Delta S=1$ decays
with Operator Product Expansion (OPE), strong FSI effects can be treated 
without introducing 
any dependence on the factorization scale which separates short--distance 
and long--distance contributions.
\end{itemize}

The last point means that the strong FSI problem in $K\to\pi\pi$ decays
can be solved independently of the matching problem between short--distance 
Wilson coefficients and long--distance weak matrix elements. 
The situation can be different for higher energy processes like 
rare $B$ decays such as $B\to\pi\pi$, where, at the B meson mass, hard 
rescattering and soft inelastic rescattering contributions are expected to be 
the dominant strong FSI effects. 
 
In Section~\ref{OMNES} the general Omn\`es problem is formulated, while in 
Section~\ref{KAON} I review
 the Omn\`es solution for the $K\to \pi\pi$ amplitudes \cite{FUTURE_PP}, 
mainly in the CP conserving sector. 
A few results for the CP violating amplitudes, relevant 
for the prediction of  $\varepsilon'/\varepsilon$ are also discussed. 
A new Standard Model prediction of $\varepsilon'/\varepsilon$ 
with the inclusion of FSI effects is discussed in \cite{SCIMEMI}. 

\section{The Omn\`es problem}
\label{OMNES}

Let us consider a generic amplitude (or form factor) $A^I_J(s)$,
with two pions in the final state
which have total angular momentum and isospin given by $J$ and $I$,
respectively, and invariant mass 
$s\equiv q^2\equiv \left(p_1 + p_2\right)^2$.
The amplitude is analytic everywhere except for a cut on the real positive 
$s$ axis $L = [4 m_\pi^2, \infty )$. 

Below the first inelastic threshold, only the $2\pi$ intermediate state 
contributes to the absorptive part of the amplitude and Watson's theorem 
\cite{Watson} implies that the phase of the amplitude is equal to the  
phase of the $\pi\pi$ partial--wave
scattering amplitude, so that
\ba\label{eq:WASIM_EL}
&&\hspace{-0.6cm}\mbox{Im}\, A^I_J \, =\, (\mbox{Im}\, A^I_J )_{2\pi} = 
 e^{-i\delta^I_J} \,\sin{\delta^I_J}\; A^I_J  \, =\, \nonumber\\
&&\hspace{-0.6cm}  e^{i\delta^I_J} \,\sin{\delta^I_J}\; A^{I\ast}_J  = 
\sin{\delta^I_J}\; \vert A^I_J\vert \, =\, \tan{\delta^I_J}\; 
\mbox{Re}\, A^I_J\, .
\ea
Cauchy's theorem implies instead that $A^I_J(s)$ can be written
as a dispersive integral along the physical cut:
\be\label{eq:DISP}
A^I_J(s) = {1\over \pi}\int_{L}~ d z \; 
{{\mbox{Im}}\, A^I_J(s)\over z-s-i\epsilon} 
\, +\, {\mbox{subtractions}}\, .
\ee
Inserting eq. (\ref{eq:WASIM_EL}) in the dispersion relation 
(\ref{eq:DISP}), one obtains an integral equation for $A^I_J(s)$ of 
the Omn\`es type, 
which has the well--known Omn\`es solution \cite{ALLOMNES}
(for $n$ subtractions with subtraction point $s_0$ outside the physical cut):
\be\label{eq:NSUB} 
A^I_J(s) \, = \, Q^I_{J,n}(s,s_0)~\exp{\left\{I^I_{J,n}(s,s_0)\right\}}
\, ,
\ee
where
\ba\label{eq:In}
&&\hspace{-0.6cm}I^I_{J,n}(s,s_0)\, \equiv\,  \nonumber\\
&&\hspace{-0.6cm} {(s-s_0)^{n} \over \pi}\,
\int^\infty_{4m^2_\pi} \,
{dz\over (z-s_0)^{n}}\,
{\delta^I_J(z) \over z-s-i\epsilon}  
\ea
and 
\ba\label{eq:NSUB_Q} 
&&\hspace{-0.6cm}\log{\left\{Q^I_{J,n}(s,s_0)\right\}}\, \equiv\, \nonumber\\
&&\hspace{-0.6cm} \sum_{k=0}^{n-1} 
{ (s-s_0)^{k} \over k!}\; {d^k \over d s^k}   
\left.\ln{\left\{A^I_J(s)\right\}} \right|_{s=s_0} ,
\ea
for $(n\geq 1)$ and with $Q^I_{J,0}(s,s_0)\equiv 1$.
The dispersive integral $I^I_{J,n}(s,s_0)$ 
is uniquely determined up to a polynomial ambiguity
(that does not produce any imaginary part of the amplitude),
which depends on the number of subtractions and the subtraction point.
The simple iterative relation for the real part of $I^I_{J,n}(s,s_0)$
\ba\label{eq:ITER}
&&\hspace{-0.6cm} \mbox{Re}\,  I^I_{J,n}(s,s_0) \, =\, 
\mbox{Re}\,  I^I_{J,n-1} (s,s_0) \nonumber\\ 
&&\hspace{0.6cm} - (s-s_0)^{n-1}\lim_{s\to s_0} {\mbox{Re}\, I^I_{J,n-1}(s,s_0)
\over (s-s_0)^{n-1}}
\, ,
\ea
shows that only a polynomial part of  $I^I_{J,n}(s,s_0)$ does depend on the 
subtraction point $s_0$ and the number of subtractions $n$, while 
the non--polynomial part of  $I^I_{J,n}(s,s_0)$,
the one containing the infrared chiral logarithms, is universal (i.e. 
$s_0$ and $n$ independent). 
Thus, the Omn\`es solution predicts the chiral logarithmic corrections
in a universal way and 
provides their exponentiation to all orders in the chiral expansion.
The polynomial ambiguity of $I^I_{J,n}(s,s_0)$ and the subtraction 
function  $Q^I_{J,n}(s,s_0)$ can be fixed, at a given order in the 
chiral expansion, by matching the Omn\`es formula (\ref{eq:NSUB}) 
with the ChPT prediction of $A^I_J(s)$. It remains a polynomial ambiguity at 
higher orders.
Notice that in the presence of a zero of the amplitude the Omn\`es solution 
can be found for the factorized amplitude  $\overline{A^I_J}(s)$, such that 
 $A^I_J(s) = (s-\zeta)^p \,\overline{A^I_J}(s)$, where $\zeta$ is a zero of 
order $p$.

\section{$K\to\pi\pi$ matrix elements}
\label{KAON}

The usual OPE description of $K\to\pi\pi$ decays is realized
 by a a three-flavour 
short--distance effective Lagrangian with $\Delta S=1$ \cite{GW:79,BURAS}, 
\be\label{eq:Leff} 
 {\cal L}_{\mathrm eff}^{\Delta S=1}= - \frac{G_F}{\sqrt{2}} 
 V_{ud}^{\phantom{*}}\,V^*_{us}\,  \sum_i  C_i(\nu) \; Q_i (\nu) \; ,  
 \label{eq:lag}  
\ee 
where $G_F$ is the Fermi coupling and $V_{ij}$ are the appropriate CKM 
matrix elements. The sum is over the product of local four--fermion 
operators $Q_i$
and the short--distance Wilson coefficients $C_i(\nu)$.
The renormalization (or factorization) scale $\nu$ separates the short--
and long--distance contributions contained in   $C_i(\nu)$ and $Q_i$ 
respectively.
The long--distance realization of matrix elements among
light pseudoscalar mesons such as $K\to\pi\pi$ can be realized 
with ChPT, as an expansion in powers of momenta of the external particles
and light quark masses.  
At lowest order in the chiral expansion, 
the most general effective bosonic Lagrangian, with the same 
$SU(3)_L\otimes SU(3)_R$ transformation properties as the short--distance 
Lagrangian (\ref{eq:lag}), contains three terms \cite{FUTURE_PP}: 
\ba\label{eq:lg8_g27} 
\cL_2^{\Delta S=1} &=& -{G_F \over \sqrt{2}}  V_{ud}^{\phantom{*}} V_{us}^* 
\Bigg\{ g_8\; f^4\;\langle\lambda L_{\mu} L^{\mu}\rangle  
 \nonumber\\
&& + g_{27}\; f^4
 \,\left( L_{\mu 23} L^\mu_{11} + {2\over 3} L_{\mu 21} L^\mu_{13} 
\right)   \nonumber\\ 
&& + e^2 f^6 g_{EM} \;  
\langle\lambda U^\dagger \cQ U\rangle \Bigg\} + 
\mbox{\rm h.c.} \, .  
\ea 
The flavour--matrix operator $L_{\mu}=-i U^\dagger D_\mu U$  
represents the octet of $V-A$ currents at lowest order in derivatives,
where $U = \exp{(i\sqrt{2}\phi/f)}$ is the exponential representation of the 
light pseudoscalar meson field with $\phi$ the flavour octet matrix.
$\cQ= {\rm diag}(\frac{2}{3},-\frac{1}{3},-\frac{1}{3})$ is the quark 
charge matrix, 
$\lambda\equiv (\lambda^6 - i \lambda^7)/2$ projects onto the 
$\bar s\to \bar d$ transition [$\lambda_{ij} = \delta_{i3}\delta_{j2}$]
and $\langle A\rangle $ denotes the flavour trace of $A$.

At generic values of the squared 
centre--of--mass energy $s=(p_{\pi 1}+p_{\pi 2})^2$, 
the $I=0,2$ amplitudes generated by the lowest--order lagrangian in eq.
(\ref{eq:lg8_g27}) are given by
\ba
&&\hspace{-0.6cm} A_0(s)\, =\, 
-{G_F\over \sqrt{2}} V_{ud}V^\ast_{us}\,\sqrt{2} f\,
 \nonumber\\
&&
\left\{\left(g_8+{1\over 9}\, g_{27}\right) (s-M_\pi^2) 
 -{2\over 3} f^2 e^2 g_{EM}\right\}  ,
\nonumber\\
&&\hspace{-0.6cm} A_2(s)\, =\,
  -{G_F\over \sqrt{2}} V_{ud}V^\ast_{us}\, {2\over 9} f
\, \left\{5\, g_{27}\, (s-M_\pi^2)\, \right .\nonumber\\
&&\left .- 3 f^2 e^2 g_{EM}\right\} 
\label{TREE}
\ea
where the usual isospin decomposition
\ba\label{eq:AI} 
A[K^0\to \pi^+\pi^-] &\equiv& {\cal A}_0 + {1\over \sqrt{2}}\, {\cal A}_2 
\, , \nonumber\\ 
A[K^0\to \pi^0\pi^0] &\equiv&  {\cal A}_0 - \sqrt{2}\, {\cal A}_2\, , 
\\ 
A[K^+\to \pi^+\pi^0] &\equiv&  {3\over 2}\, {\cal A}_2 \, , \nonumber 
\ea 
has been used. In the absence of $e^2 g_{EM}$ corrections,
the amplitudes in eq. (\ref{TREE}) have a zero at $s=M_\pi^2$,
because the on-shell amplitudes should vanish in the SU(3) limit
\cite{SU3}. This is not the case for the amplitudes mediated by the 
electroweak penguin operator $Q_8$, since its lowest--order
ChPT realization is given by the term proportional to $e^2 g_{EM}$.
The lowest order chiral contribution to those amplitudes is a constant
of order $e^2 p^0 $ (which anyway counts as $O(p^2)$ in the usual chiral 
power counting). 

The derivation of the Omn\`es solution for $K\to\pi\pi$ decays has been 
discussed in detail in ref. \cite{FUTURE_PP}. Here, I focus on some relevant
aspects of the problem. Our aim is to resum the strong FSI effects 
due to soft rescattering of the two pions in the final state.
The study of the scalar pion form factor and its comparison with $K\to\pi\pi$
 amplitudes in ref. \cite{FUTURE_PP} has clarified various facts:
\begin{itemize}
\item[$\bullet$] 
soft FSI in the $I=0$ channel generate large infrared logarithms dependent on 
the pion mass which need to be resummed to all orders in ChPT.
\item[$\bullet$] 
Those infrared logarithms are universal, i.e. only depend on the quantum 
numbers of the $\pi\pi$ system in the final state.
\item[$\bullet$]
The Omn\`es solution provides an evolution 
of the given amplitude from low energy values, where the ChPT momentum 
expansion can be trusted, to higher energy values, through the exponentiation 
of the infrared effects due to FSI.
 \end{itemize}
However, one difference between the Omn\`es solution for $K\to\pi\pi$ 
amplitudes and the scalar pion form factor, is that we need to consider 
an off-shell kaon of mass squared  $s=(p_{\pi 1}+p_{\pi 2})^2$ in the first 
case, instead of a physical momentum transfer $s$. 
This generates a {\em local} ambiguity at higher orders in the ChPT expansion
(see also \cite{FUTURE_PP} for the explicit expressions at the 
next--to--leading order in ChPT), which however has nothing to do with 
the Omn\`es procedure of resumming FSI effects.

The CP conserving $K\to\pi\pi$ isospin amplitudes admit a general 
decomposition in ChPT \cite{FUTURE_PP}:
\be\label{eq:split} 
\cA_I(s) = \tilde{a}_I(s) \left(s-M_\pi^2\right) + 
\delta \tilde{a}_I(s) \left(M_K^2-M_\pi^2\right)\, , 
\ee 
where $\delta \tilde{a}_I(s)$ parameterizes tiny corrections due to the 
explicit breaking of chiral symmetry via the quark mass matrix and it 
is zero at lowest order\footnote{ 
%%%%%
To make the decomposition \protect(\ref{eq:split}) unique,
we require the function $\delta\tilde{a}_I(s)$ to depend on $s$ only 
logarithmically. }. 
%%%%% 
Since there is a single strong phase, for a given isospin, 
the unitarity relation (\ref{eq:WASIM_EL}) is valid for
 $\tilde{a}_I(s)$ and $\delta\tilde{a}_I(s)$ individually and  
 the Omn\`es problem can be solved separately for the 
two pieces. 
Combining them, one can write the result for the physical 
on-shell amplitude in the simpler form: 
\ba\label{eq:OMNES_WA} 
&&\hspace{-0.6cm}\cA_I \,\equiv \,\cA_I(M_K^2) =   
\left(M_K^2-M_\pi^2\right) \; a_I(M_K^2) \nonumber\\
&&\, =\,   
\left(M_K^2-M_\pi^2\right) \; \Omega_I(M_K^2,s_0) \; a_I(s_0)\\
&&\, = \,
\left(M_K^2-M_\pi^2\right) \; \Re_I(M_K^2,s_0) \; a_I(s_0) 
\; e^{i\delta^I_0(M_K^2)}\, , 
\nonumber\ea 
where $a_I(s)\equiv \tilde{a}_I(s) + \delta\tilde{a}_I(s)$.
The Omn\`es factor $\Omega_I(M_K^2,s_0)$ can be interpreted as a sort of
evolution operator from the subtraction point $s_0$ to $M_K^2$. 
Its explicit expression for a given number of subtractions can be directly 
derived from eq. (\ref{eq:NSUB}) and can be split into the dispersive 
contribution $\Re_I(M_K^2,s_0) $ and the phase shift exponential. 
Notice also that the once--subtracted Omn\`es factor 
$\Omega_I^{(1)}(M_K^2,s_0)$ is universal because it only depends on the 
phase shifts $\delta^I_0(s)$, while for two subtractions
the Omn\`es factor depends on $f'(s_0)/f(s_0)$ for a given amplitude $f(s)$.
However, given the smallness of the sub-leading $\delta\tilde{a}_I$ 
contribution, it remains a good numerical approximation to take a global 
Omn\`es factor for $a_I(s)$ also with two subtractions.
 
%In the numerical analysis a subtraction point $s_0 =0$ has been chosen
% \cite{FUTURE_PP} 
%in order to minimize the contribution from unknown higher order ChPT 
%corrections to $a_I(s_0)$. 
For each of the amplitudes $a_0^{(8)}$, 
$a_0^{(27)}$ (the octet and 27-plet $I=0$ amplitudes) and $a_2$ (with $I=2$),
the $s$ dependence can be written in a simple form:
\be\label{eq:s_dep} 
a(s) = a(0)\;\left\{ 1 + g(s) 
  + O(p^4)\right\}\, , 
\ee
where the one--loop functions  $g = g_0^{(8)}$, $g_0^{(27)}$ and $g_2$
have been computed in ref. \cite{FUTURE_PP}. The main properties of the 
$g(s)$ functions can be summarized as follows:
\begin{itemize}
\item[$\bullet$]
 the contribution from  $ \delta\tilde{a}_I(s)$ is always very small 
and exclusively due to non--analytic $\bar{K}K$ and $\eta\eta$ loop 
contributions which are numerically suppressed at low values of $s$.
 \item[$\bullet$] All isoscalar $g$ functions contain exactly the same 
infrared  $\ln{M_\pi^2}$ contribution and the same contribution from the 
finite one--loop $\pi\pi$ rescattering function $\bar{J}_{\pi\pi}(s)$ 
\cite{FUTURE_PP} which generates the absorptive part of the isospin 
amplitude below the inelastic threshold. This shows the universality 
of the infrared effects due to FSI.
\item[$\bullet$]
 The $s$ dependence of the one--loop correction at low values of $s$ 
is dominated by the pure $SU(2)$ effect of elastic $\pi\pi\to\pi\pi$ 
scattering. These universal infrared effects enhance the $I=0$ amplitudes 
while suppress the $I=2$ amplitude. 
\end{itemize}
The dynamics leading to the $\pi\pi$ final state, also generates local 
contributions  which are different in each case. For the scalar form factor 
these contributions are small \cite{FUTURE_PP}. 
For the weak $K\to\pi\pi$ amplitudes 
the knowledge of those contributions (generated by the ChPT counterterms)
is still quite limited and has to be further investigated. 
In addition, being the kaon off-shell, local off-shell contributions 
 are also allowed, starting at next--to--leading order in the chiral 
expansion.
 The usual factorization models \cite{EKW:93} predict all the local
 contributions to the functions $g(s)$ to be zero at the ChPT 
renormalization scale $\mu = M_\rho$.
However, a model--independent analysis still remains affected by 
the ambiguity due to the presence of {\em local} contributions, already 
for the on--shell amplitude.  The Omn\`es factor cannot fix 
that problem. The role of the Omn\`es factor remains that of providing an
efficient resummation of large infrared effects due to FSI. The advantage
of the Omn\`es exponentiation respect to the usual one--loop ChPT 
computation is to control the uncertainty coming from higher order
($\geq$ two--loops) FSI effects.

Taking a low subtraction point $s_0 =0$ where higher--order 
corrections are expected to be small, we can just multiply the tree--level 
formulae (\ref{TREE}) with the experimentally determined Omn\`es 
exponentials \cite{FUTURE_PP}.
The two dispersive corrections factors thus obtained \cite{FUTURE_PP} are 
$\Re_0(M_K^2,0)\,=\, 1.55 \pm 0.10$  and 
$\Re_2(M_K^2,0) \,=\, 0.92 \pm 0.03$, 
where the errors are supposed to take into account a) the uncertainties of the 
fits to the 
experimental phase shifts data used in the calculation of the Omn\`es factor
and b) 
the additional inelastic contributions above the first inelastic threshold.

The corrections induced by FSI in the moduli of the decay amplitudes 
${\cal A}_I$ generate an additional enhancement of the 
$\Delta I=1/2$ to $\Delta I=3/2$ ratio, 
\be\Re_0(M_K^2,0)/\Re_2(M_K^2,0) = 1.68\pm 0.12 \, .\label{eq:ratio}\ee 
This factor multiplies the enhancement already found at short distances. 

The Omn\`es procedure can be directly extended to the CP--violating 
$K\to\pi\pi$ amplitudes relevant for the estimate of the direct CP violation 
parameter $\varepsilon^\prime /\varepsilon$. 
In deriving the structure of the absorptive part of 
the amplitude in eq. (\ref{eq:WASIM_EL}) one makes use of Time--Reversal 
invariance, so that the Omn\`es procedure as formulated in eq. (\ref{eq:NSUB})
can be applied only to CP--conserving amplitudes.
However, working at the first order in the Fermi coupling, the CP--odd phase 
is fully contained in the ratio of CKM matrix elements 
$\tau = V_{td}\,V^*_{ts} / V_{ud}\,V^*_{us} $ which multiplies 
the short--distance Wilson coefficients. Decomposing the isospin amplitude 
as $\cA_I = \cA_I^{CP} + \tau \,\cA_I^{\slashchar{CP}} $, the Omn\`es 
solution can be derived for the two amplitudes $ \cA_I^{CP} $ and
 $\cA_I^{\slashchar{CP}}$ which respect Time--Reversal invariance.
In ref. \cite{SHORT} it has been shown how the inclusion of FSI effects 
in $K\to\pi\pi$ amplitudes can easily enhance  
 previous short--distance based Standard Model predictions of 
$\varepsilon^\prime /\varepsilon$ \cite{Buras,Lattice}
 by roughly a factor of two.  
To obtain a complete Standard Model prediction for 
$\varepsilon^\prime /\varepsilon$ an exact matching procedure has been
 proposed \cite{FUTURE_PPS}.
It is  inspired 
by the large--$N_c$ expansion, but only at scales below the charm 
quark mass $\mu \leq m_c$ (where the logarithms that enter the Wilson 
coefficients are small). FSI effects, which are next--to--leading 
in the $1/N_c$ expansion but numerically relevant, are taken into 
account through the multiplicative factors $\Re_I(M_K^2,0)$ while 
avoiding any double counting. The Standard Model prediction for 
$\varepsilon^\prime /\varepsilon$ has been discussed in \cite{SCIMEMI} 
at this conference.

\section*{Acknowledgments}
I warmly thank my collaborators Antonio Pich and Ignazio Scimemi
and the organizers of the conference.
 This work has been supported  by the Ministerio de Educaci\'on  y Cultura 
(Spain) and in part by the European Union TMR Network EURODAPHNE 
(Contract No. ERBFMX-CT98-0169).

%\section*{References}

\end{document}